%% file: paper-na64.tex
\RequirePackage{lineno} 
\documentclass[aps,prl,twocolumn,showpacs,groupedaddress,superscriptaddress,amsmath,amssymb]{revtex4-1}

\usepackage{graphicx}
\usepackage{dcolumn}
\usepackage{bm}
\usepackage{amssymb}
\usepackage{enumerate}
\usepackage{lineno}

\newcommand\g{\gamma}
\newcommand\ee{e^+e^-}
\newcommand\pp{\pi^+ \pi^-}
\newcommand\mm{\mu^+ \mu^-}
\newcommand\xee{X \to e^+e^-}
\newcommand\aee{A' \to e^+e^-}
\newcommand\xdecay{X \rightarrow e^+  e^-}

\newcommand\ma{m_{A'}}
\newcommand\pair{e^+e^-}

\newcommand\Na{{N}_{A'}}

\newcommand\kosdec{K^0_S \to \pi^0 \pi^0; \pi^0 \to \g \ee}

\newcommand\kospi{K^0_S \to \pi^+ \pi^-}
\newcommand\ks{K^0_S }

\def\address{\@ifstar{\address@star}%
  {\@ifnextchar[{\address@optarg}{\address@noptarg}}}

\begin{document}

\title{ Search for a Hypothetical 16.7 MeV Gauge Boson and Dark Photons \\
in the NA64 Experiment at CERN}
\input author_list.tex

\date{\today}

\begin{abstract}
  We report the first results on a direct search for a new 16.7 MeV boson ($X$) which could explain the anomalous excess of $\ee$ pairs 
observed in the excited $^8$Be$^*$ nucleus decays. Due to its coupling to electrons, the X could be produced in the bremsstrahlung
reaction $e^- Z \to e^- Z  X$ by a 100 GeV $e^-$ beam  incident on an active target in the NA64 experiment at the CERN SPS and 
observed through the subsequent decay into a $\ee$ pair.
With $5.4\times 10^{10}$ electrons on target no evidence for such decays was found, allowing to set first limits on the $X-e^-$
coupling in the range $ 1.3\times 10^{-4}\lesssim \epsilon_e \lesssim 4.2\times 10^{-4}$ excluding part of the allowed
parameter space. We also set new bounds on the mixing strength of photons with dark photons ($A'$) from non-observation of
the decay $\aee$ of the bremsstrahlung $A'$ with a mass $\lesssim 23$ MeV.
\end{abstract}
\maketitle

 The ATOMKI experiment of Krasznahorkay et al. \cite{be8} has reported the observation of a 6.8 $\sigma$ excess of events
in the invariant mass distributions of $\pair$ pairs produced in the nuclear transitions of excited  $^8Be^*$ to its ground state via
internal pair creation.  This anomaly can be interpreted as the emission of a new 
protophobic gauge $X$ boson with a mass of 16.7 MeV followed by its $\xdecay$ decay assuming that the $X$ has
non-universal coupling to quarks, coupling
to electrons in the range $2\times 10^{-4} \lesssim \epsilon_e \lesssim 1.4\times 10^{-3}$ and the lifetime 
$10^{-14}\lesssim \tau_X \lesssim 10^{-12}$~s \cite{feng1,feng2}. 
It has motivated worldwide theoretical and experimental 
efforts towards light  and weakly coupled  vector bosons,  see, e.g. \cite{mb,  nardi, jk, cheng, Zhang:2017zap, ia,
liang, bart, pf}. \par Another strong motivation to the search for a new light boson decaying into $\ee$ pair is provided by the
Dark Matter puzzle. An intriguing possibility is that in addition to gravity a new effective force between the dark sector and
visible matter, transmitted by a new vector  boson,  $A'$ (dark photon), might exist \cite{prw, pospelov}. Such $A'$ could have 
a mass $m_{A'}\lesssim 1$ GeV, associated with a spontaneously broken gauged $U(1)_D$ symmetry, and would couple to 
the Standard Model (SM) through kinetic mixing with the ordinary photon, $-\frac{1}{2}\epsilon F_{\mu\nu}A'^{\mu\nu}$, parametrized
by the mixing strength  $\epsilon \ll 1$ \cite{Okun:1982xi, Galison:1983pa, Holdom:1985ag}. For a review see, e.g. \cite{mb, jr, report}. 
\begin{figure*}[tbh!!]
\centering
\includegraphics[width=.8\textwidth]{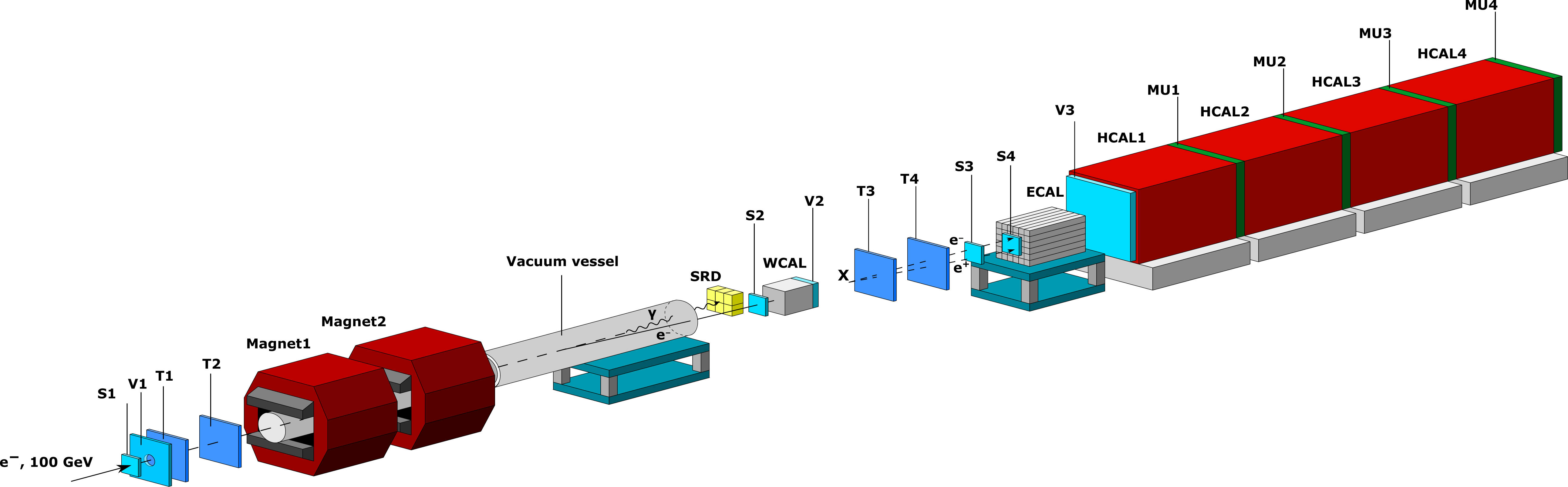}
\caption{Schematic illustration of the NA64 setup to search for  the $A',X\to \ee$  decays.}
\label{setup}
\end{figure*}
 A number of previous  beam dump \cite{jdb, charm, rio, e137, konaka, bross, dav,  ath, nomad, e787, essig1, blum,
sg1, blum1, sarah1}, fixed target \cite{apex,merkel,merkel1}, collider \cite{babar, curt, babar1} and rare particle decay
\cite{bern, sindrum, kloe, sg2, kloe2, kloe4, wasa, hades, phenix, e949, na48, pol, kloe3} experiments have already put stringent
constraints on the mass $m_{A'}$ and $\epsilon$ of such dark photons excluding, in particular,  
the parameter space  region favored by the $g_\mu-2$ anomaly. However,  the range of mixing strengths 
$10^{-4} \lesssim \epsilon \lesssim 10^{-3}$ corresponding to a short-lived $A'$ still remains unexplored. 
 In this Letter we report the first results from the NA64 experiment specifically designed for a direct
search of the $\ee$ decays of new short-lived particles  in the sub-GeV mass range at the CERN SPS \cite{Gninenko:2013rka,
Andreas:2013lya,na64-prl,na64-prd}.
 \par The method of the search for $\aee$ decays is described in \cite{Gninenko:2013rka, Andreas:2013lya}. Its  application
 to the case of the  $\xee$  decay is straightforward. Briefly, a 
  high-energy electron beam is sent into an electromagnetic (e-m) calorimeter that serves as an active beam dump.
Typically the beam electron loses all its  shower energy in the dump.  If the 
$A'$ exists, due to the $A'(X) - e^-$ coupling it would occasionally be produced    by a shower electron (or positron) 
in its scattering off a nuclei of the dump:
\begin{equation}
e^- + Z \to e^- + Z + A'(X)   ;~ A'(X)\to \ee .
\label{ea}
\end{equation}
Since the $A'$ is penetrating and longer lived, it would escape the beam dump, and
subsequently decays into an $\ee$ pair in a downstream set
of detectors. The pair energy would be equal to the energy missing
from the dump. The apparatus is designed to identify and measure
the energy of the $\ee$ pair in another calorimeter (ECAL).
Thus, the signature of the  $A'(X) \to \ee$ decay  is an event with two e-m-like showers
in the detector: one shower in the dump, and another one in the ECAL with the sum energy  equal to the beam energy.   
\par  The NA64 setup  is schematically shown in Fig.~\ref{setup}. 
The experiment  employs the optimized 100 GeV electron beam from the H4 beam line in the North Area (NA) of the CERN SPS.
Two scintillation counters, S1 and S2 were used for the beam definition, while the other two, S3 and S4,
were used to detect the $\pair$ pairs. The detector was equipped with two dipole magnets
and a tracker, which was a set of four upstream Micromegas (MM) chambers (T1, T2) for the incoming
$e^-$ angle selection and two sets of downstream MM, gas electron multiplier (GEM) stations and scintillator hodoscopes (T3, T4) for  measurements
of the outgoing tracks \cite{Banerjee:2015eno,track}.
To enhance the electron identification the synchrotron radiation (SR) emitted by electrons was used for their 
tagging allowing to  suppress  the initial hadron contamination in the  beam $\pi/e^- \simeq 10^{-2}$
down to the level $\simeq 10^{-6}$ \cite{srd,na64-prd}.
The use of SR detectors (SRD) was a key point for the improvement of the sensitivity compared 
to the previous electron beam dump searches \cite{konaka,bross}. The dump was  a compact e-m calorimeter WCAL
made as short as possible to maximize the sensitivity to short lifetimes while keeping the leakage of particles at a small level.
It was  followed by the ECAL to measure the energy of the decay $\ee$ pair, which was a matrix of $6\times 6 $ shashlik-type modules \cite{na64-prd}.
The ECAL has $\simeq 40$ radiation lengths ($X_0$) and is located at a distance $\simeq 3.5$~m from the WCAL.
Downstream of the ECAL the detector was equipped with a high-efficiency veto counter, V3, and a hermetic hadron
calorimeter (HCAL) \cite{na64-prd} used as a hadron veto and for muon identification with a help
of four muon counters, MU1-MU4, located between the HCAL modules. The results reported here  were obtained from data samples in which $2.4\times 10^{10}$ electrons on target (EOT) and $3\times 10^{10}$
EOT were collected with the WCAL of 40 $X_0$ (with a length of 290 mm) and of 30 $X_0$ (220 mm), respectively. The events were collected
with a hardware trigger requiring in-time energy deposition in the WCAL and $E_{WCAL} \lesssim 70$ GeV.
Data of these two runs (hereafter called the 40 $X_0$ and 30 $X_0$ run) were analyzed with similar selection criteria and finally
summed up, taking into account the corresponding normalization factors.
 For the mass range $1 \leq m_{A'} \leq~25$ MeV and energy $E_{A'}\gtrsim 20$ GeV, the opening angle $\Theta_{\ee} \simeq 2 m_{A'}/E_{A'} \lesssim 2$
mrad of the decay $\ee$ pair is too small to be resolved in the tracker T3-T4, and the pairs are mostly detected as a single-track
e-m  shower in the ECAL.
\par The candidate events were selected with the following criteria chosen to maximize the signal acceptance and
 minimize background, using both Geant4 \cite{Agostinelli:2002hh,geant} based simulations and data:
(i) There should be only one track entering the dump. No cuts on reconstructed outgoing tracks were used;
(ii) No  energy deposition in the V2 counter exceeding about half of the energy deposited by the  minimum ionizing particle (MIP);
(iii) The signal in the decay counter S4 is consistent with two MIPs;  
(iv) The sum of energies deposited in the WCAL and ECAL, $E_{tot} = E_{WCAL} +E_{ECAL}$,  is equal to the beam energy within the energy resolution of these detectors.
 According to simulations, at least 30\% of the total energy should be deposited in the ECAL \cite{gkkk1,gkkk2};
(v) The showers in the WCAL and ECAL should start to develop within a few first $X_0$;
(vi) The lateral and longitudinal shape of the shower in the ECAL are consistent with a single e-m one. 
However,  for  $A'$s with the energy $\lesssim 5$ GeV the ECAL shower is poorly 
described by the single shower shape, hence the additional cut $E_{ECAL} > 5$ GeV was applied;
(vii) No significant energy deposited in the V3 and/or  HCAL. These cuts were used for  rejection of events with hadrons in the final state.  
 As in the previous  analyses \cite{na64-prl, na64-prd} a clean sample of $\simeq 10^5$ rare $\mu^+ \mu^-$ events produced in the dump 
 was used for the   efficiency corrections in the simulations, which do not exceed 20\%. A blind analysis of data was performed, with 
the signal box defined as $90 < E_{tot} < 110 $ GeV and by using  20\%(100\%)  of the data for the selection criteria optimization (background estimate).
\begin{figure}
\centering
\includegraphics[width=0.45\textwidth]{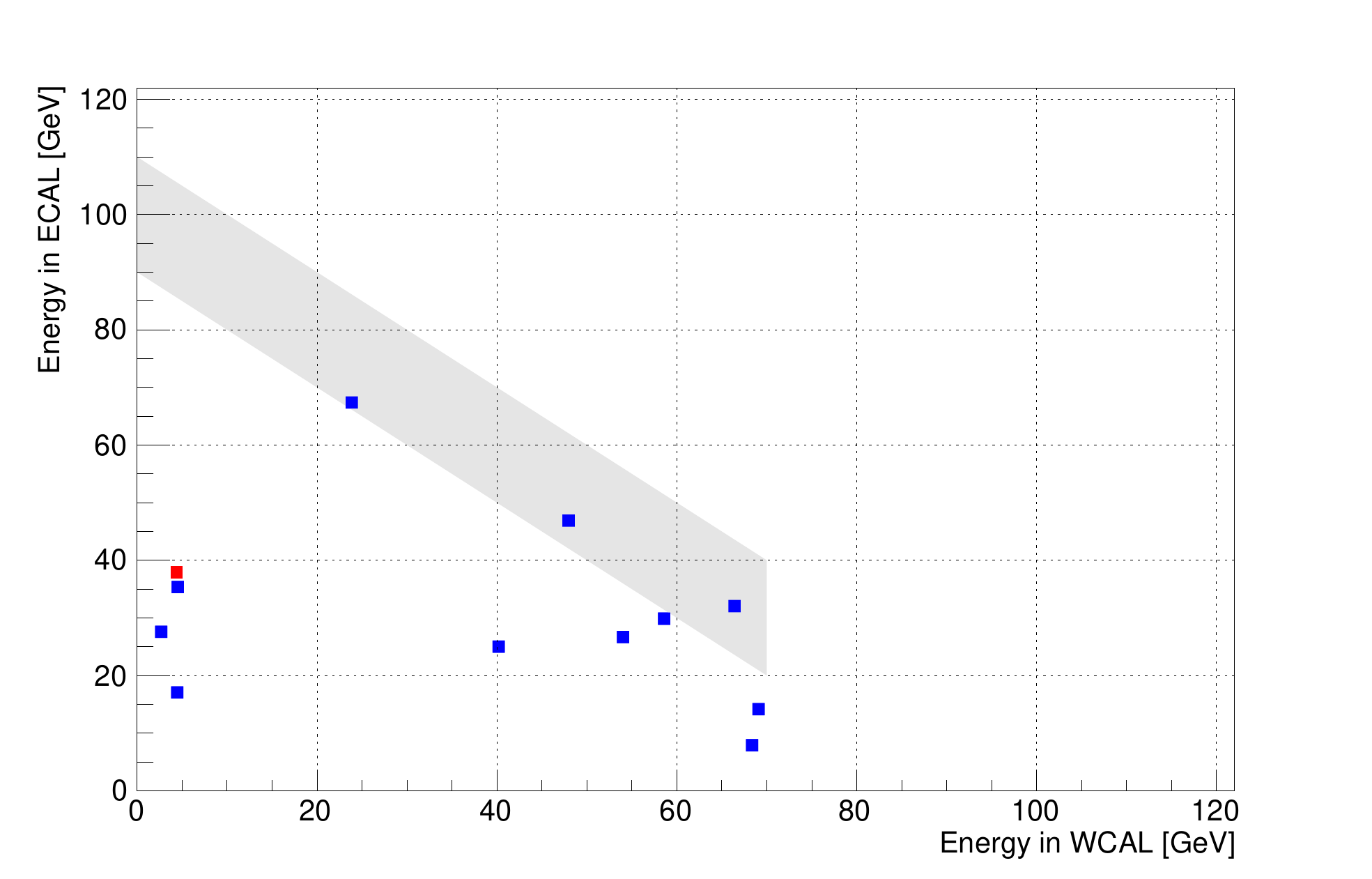}
\caption{Distribution of selected e-m neutral (presumably photon)  and signal events in the $(E_{WCAL};E_{ECAL})$ plane from the combined 30 $X_0$ and 40 $X_0$ runs. Neutral e-m events  are shown as blue squares. The only signal-like event is shown as a red square. The dashed band  represents
the signal box. }
 \label{w-e}
\end{figure}
\par There are several processes that can fake the $\aee$ signal. Among them, the two most important were expected either from decay chain 
 $\kosdec$ of $\ks$ produced in the WCAL or from the $\g\to \ee$ conversion of photons from $\ks\to \pi^0 \pi^0 \to \pi^0\to \g \g$ decays in the T3 plane or earlier in the beamline.
 Another background could come from the $\kospi$ hadronic decays that could be misidentified as an e-m event in the 
ECAL at the level $\lesssim 2.5 \times 10^{-5}$ evaluated from the measurements with the pion beam.
 The leading $K^0$ can be produced in the dump either by misidentified beam $\pi^-, K^-$  or directly by electrons.
The background from the $\ks$ decay chain was estimated by using the direct measurements of the $K^0_S$ flux from the dump 
with the following method. It is well known that the $K^0$ produced in hadronic reactions is a linear combination of the short- and
long-lived components $|K^0> = (|K^0_S> + |K^0_L>)/\sqrt{2}$. The flux of $K^0$ was evaluated from the measured ECAL+HCAL energy 
spectrum of long-lived neutral hadrons selected with the requirement of no signal in V2 and S4, taking into account
corrections due to the $K^0_S$ decays in-flight. The main fraction of $\simeq 10^3$ events observed in the HCAL 
were neutrons produced in the same processes as $K^0$ in the WCAL. According to simulations, $\lesssim 10\%$ of them were predicted
to be other neutral hadrons, i.e. $\Lambda$ and $K^0$, that were also included in the data sample. The conservative assumption that
$\simeq 100$ $K^0$ were produced allows us to calculate the number of $K^0_S$ from the dump and simulate the corresponding background
from the $\kospi$ and $\kosdec$ decay chain, which was found to be $\lesssim 0.04$ events per $5.4\times 10^{10}$ EOT.
To cross-check this result another estimate of this background was used. The true neutral e-m events, which are presumably photons,  were selected with requirements of no charged tracks, i.e. no  signals in V2 and S4 counters,  plus a single e-m like shower in the ECAL defined by cuts  cuts (v)-(vii). 
Three such events were  found in the signal box as
shown in Fig. \ref{w-e}. Using simulations we calculated that there were $\simeq 150$  leading $K^0$  produced in the dump, which is 
in a reasonable agreement with the previous estimate resulting in a conservative  $\ks$ background of 0.06 events. 
The $\mu$, $\pi$ and $K$  mistakenly tagged as $e^-$s \cite{srd} could also interact in the dump though the $\mu Z \to \mu Z \g$ or $\pi, K$
charge-exchange reactions, accompanied by the poorly detected scattered $\mu$, or secondary hadrons. 
\begin{table}[tbh!!] 
\caption{Expected numbers of background events in the signal box  estimated for
 $5.4\times 10^{10}$ EOT.}\label{tab:table2}
\begin{tabular}{lr}
\hline
\hline
Source of background& Events\\
\hline
$\ee$  pair production  by  punchthrough $\g$ & $<0.001$\\
$K^0_S\to 2\pi^0; \pi^0\to \g \ee$;$\g\to \ee$; $K^0_S\to \pi^+\pi^-$ & $0.06\pm0.034$\\
$\pi N\to (\geq 1) \pi^0 +n+...;   \pi^0\to \g \ee$;$\g\to \ee$ & $0.01\pm0.004$\\
$\pi^-$  bremsstrahlung in the WCAL , $\g\to \ee$&$<0.0001$\\
$\pi, K \to e \nu$, $K_{e4}$ decays  & $ < 0.001$ \\
$eZ\to eZ \mm; \mu^\pm \to e^\pm \nu \nu$ &  $ < 0.001$\\
punchthrough $\pi$ & $<0.003$\\
\hline 
Total    &          $0.07\pm 0.035$ \\
\hline
\hline
\end{tabular}
\end{table}
The misidentified pion could mimic the signal either directly (small fraction of showers that look like an e-m one) or by emitting a hard
bremsstrahlung photon in the last layer of the dump, which then produces an e-m- shower in the ECAL, accompanied by the scattered pion
track. Another background can appear from the beam $\pi \to e \nu$ decays downstream  the WCAL. The latter two backgrounds can  pass
the selection only due to the V2 inefficiency ($\simeq 10^{-4}$), which makes them negligible.
The charge-exchange reaction $\pi^- p \to (\geq 1) \pi^0 + n + ...$ which can occur in the last layers of the WCAL with decay
photons escaping the dump without interactions and accompanied by poorly detected secondaries is another source of fake signal.
To evaluate this background we used the extrapolation of the charge-exchange cross sections, $\sigma \sim Z^{2/3}$, measured on
different nuclei \cite{vnb}. 
The contribution from the beam kaon decays in-flight $K^-  \to e^- \nu \pp (K_{e4})$ and dimuon production in the dump 
$e^- Z \to e^- Z \mu^+ \mu^-$ with either $\pp$ or $\mu^+ \mu^-$ pairs misidentified as e-m event in the ECAL was found to be negligible. 
\par Table I summarizes the conservatively estimated background inside the signal box, which is expected to be $0.07\pm0.034$ events per
$5.4\times 10^{10}$ EOT. The dominant contribution to background is $0.06$ events from the $\ks$ decays, with  the uncertainty
dominated by the statistical error. 
 In  Fig.~\ref{w-e} the final distributions of e.m. neutral events, which are presumably photons, and signal candidate events  that passed the selection criteria (i)-(iii) and  (v)-(vii) are  shown  in the  $(E_{ECAL}; E_{WCAL})$ plane. No candidates are found in the signal box. 
The conclusion that the background is small is confirmed by the data. 
 \par The combined 90\% confidence level (C.L.) upper limits for the mixing strength $\epsilon$ were obtained  from the corresponding 
  limit for the expected number of signal events, $N_{A'}^{90\%}$,  by using the modified frequentist approach, 
taking the profile likelihood as a test statistic \cite{junk,limit,Read:2002hq}. The $N_{A'}$ value  is given by  the sum :
\begin{equation}
\Na = \sum_{i=1}^{2} N_{A'}^i = \sum_{i=1}^{2} n_{EOT}^i  \epsilon_{tot}^i n_{A'}^i(\epsilon,\ma)
\label{nev}
\end{equation}
where $\epsilon_{tot}^i$ is the signal efficiency in the run {\it i} (30 $X_0$ or 40 $X_0$), and  $n_{A'}^i(\epsilon,\ma)$  is the number  of the $\aee$ decays in the decay volume  with energy $E_{A'} > 30$ GeV per EOT,  calculated under assumption that this decay mode  is predominant, see e.g. Eq.(3.7) in Ref. \cite{Andreas:2013lya}.
Each $\it i$-th entry in this sum was calculated by simulating signal events for the corresponding beam running conditions and processing
them through the reconstruction program  with the same selection criteria and  efficiency corrections as for the data sample from the
run-{\it i}.
 \begin{figure}[tbh!!] 
\begin{center}
\includegraphics[width=0.45\textwidth]{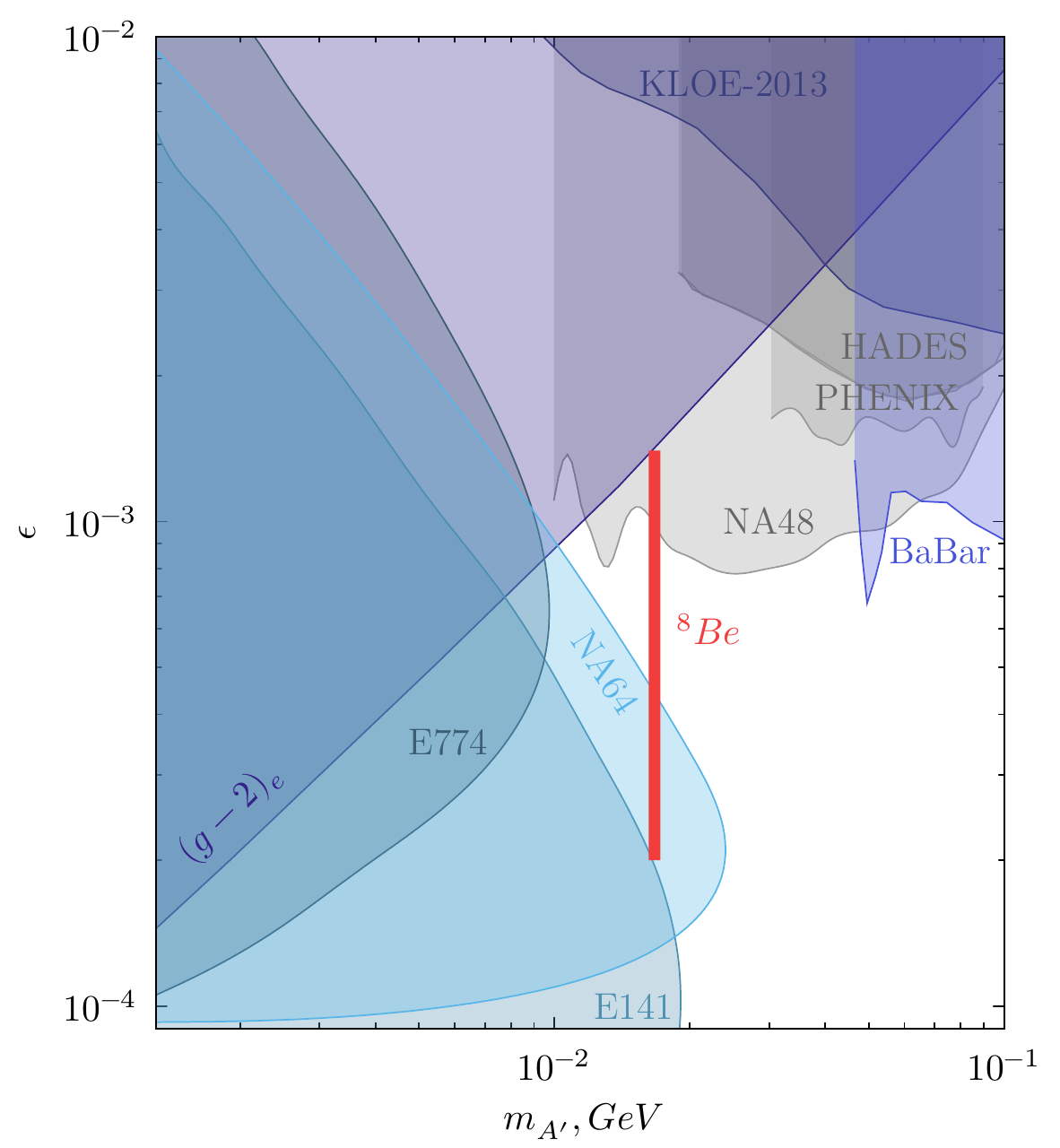}
\caption{The  90\% C.L.\ exclusion areas  in the ($m_{X}; \epsilon$) plane from the NA64 experiment (blue area). For the mass of 16.7~MeV, 
 the  $X-e^-$ coupling region excluded by NA64  is  $1.3\times 10^{-4}< \epsilon_e < 4.2~\times~10^{-4}$. The allowed range of $\epsilon_e$ 
 explaining the $^8$Be* anomaly  (red area) \cite{feng1, feng2}, constraints  on the mixing $\epsilon$ from the   experiments E141~\cite{rio},  E774~\cite{bross},  BaBar ~\cite{babar1}, KLOE~\cite{kloe2}, HADES~\cite{hades},  PHENIX~\cite{phenix}, NA48~\cite{na48},    and bounds from the electron anomalous magnetic moment $(g-2)_e$ \cite{hd}  are also shown.}
 \label{exclvis}
\end{center}
\end{figure}
The $A'$ efficiency  and its  systematic error  were determined to stem from
the overall normalization, $A'$ yield and decay probability, which were the $A'$ mass dependent, and  also  from  
efficiencies and their  uncertainties 
in the primary $e^-(0.85\pm0.02)$, WCAL($0.93\pm0.05$), $V_2(0.96\pm0.03)$, ECAL($0.93\pm0.05$), $V_3(0.95\pm0.04)$, and HCAL($0.98\pm0.02$)  event detection. The later,  shown as an example values for  the 40 $X_0$ run, were determined from measurements with $e^-$ beam  cross-checked with simulations.
A detailed simulation of the e-m shower in the dump \cite{gkkk1} with $A'$ cross sections was
used to calculate the $A'$ yield \cite{gkkk2,Liu:2016mqv,Liu:2017htz}. The $\lesssim 10\%$ difference between the calculations in Ref.\cite{gkkk2} and 
Ref.\cite{Liu:2016mqv,Liu:2017htz} was accounted for as a systematic uncertainty in $n_{A'}(\epsilon, \ma)$.  
 In the overall signal efficiency for each run the acceptance loss due to pileup ($\simeq 7\%$ for 40 $X_0$ and $\simeq 10\%$
for 30 $X_0$ runs) was taken into account and cross-checked using reconstructed dimuon events \cite{na64-prd}.
 The dimuon efficiency  corrections ($\lesssim 20\%$) were obtained with uncertainty
of 10\% and 15\%, for the 40 $X_0$ and  30 $X_0$ runs, respectively. 
 The total systematic uncertainty on $N_{A'}$ calculated by adding all errors in quadrature did not exeed $25\%$ for both runs.
 The combined 90\% C.L. exclusion limits on the mixing  $\epsilon$ as a function of the $A'$ mass is shown in Fig.~\ref{exclvis} together 
with the current constraints from other experiments. Our results exclude X-boson as an explanation for the $^8$Be* anomaly
for the $X-e^-$ coupling  $ \epsilon_e  \lesssim 4.2 \times 10^{-4}$ and mass value of 16.7 MeV, leaving the still unexplored  region
$4.2~\times  10^{-4} \lesssim  \epsilon_e  \lesssim 1.4 \times 10^{-3}$ as  quite an exciting prospect for further searches.
 \par We gratefully acknowledge the support of the CERN management and staff 
and the technical staffs of the participating institutions for their vital contributions. We also thank Attila Krasznahorkay for useful
discussions about the Atomki experiment. 
This work was supported by the HISKP, University of Bonn (Germany), JINR (Dubna), MON and RAS (Russia), 
SNSF grant 169133 and ETHZ (Switzerland), and  grants FONDECYT 1140471 and 1150792, Ring ACT1406 and Basal FB0821 CONICYT (Chile).
Part of the work on simulations was supported by the RSF grant 14-12-01430.  

\end{document}

%% file: author_list.tex
\affiliation{\it Universit\"at Bonn, Helmholtz-Institut f\"ur Strahlen-und Kernphysik, 53115 Bonn, Germany} 
\affiliation{\it  Joint Institute for Nuclear Research, 141980 Dubna, Russia}
\affiliation{\it CERN, European Organization for Nuclear Research, CH-1211 Geneva, Switzerland}
\affiliation{\it University of Illinois at Urbana Champaign, Urbana, 61801-3080 Illinois, USA}
\affiliation{\it Institute for Nuclear Research, 117312 Moscow, Russia}
\affiliation{\it P.N. Lebedev Physics Institute, Moscow, Russia, 119 991 Moscow, Russia}
\affiliation{\it Skobeltsyn Institute of Nuclear Physics, Lomonosov Moscow State University, 119991  Moscow, Russia}
\affiliation{\it Technische Universit\"at M\"unchen, Physik  Department, 85748 Garching, Germany}
\affiliation{\it Physics Department, University of Patras, 265 04 Patras, Greece} 
\affiliation{\it State Scientific Center of the Russian Federation Institute for High Energy Physics of National Research Center 'Kurchatov Institute' (IHEP), 142281 Protvino, Russia}
\affiliation{\it Tomsk State Pedagogical University, 634061 Tomsk, Russia}
\affiliation{\it Universidad T\'{e}cnica Federico Santa Mar\'{i}a, 2390123 Valpara\'{i}so, Chile}
\affiliation{\it ETH Z\"urich, Institute for Particle Physics and Astrophysics, CH-8093 Z\"urich, Switzerland}
\author{D.~Banerjee}\affiliation{\it University of Illinois, Urbana Champaign, Illinois, USA}
\author{V.~E.~Burtsev}\affiliation{\it Tomsk State Pedagogical University, 634061 Tomsk, Russia}
\author{A.~G.~Chumakov}\affiliation{\it Tomsk State Pedagogical University, 634061 Tomsk, Russia}
\author{D.~Cooke}\affiliation{\it ETH Z\"urich, Institute for Particle Physics and Astrophysics, CH-8093 Z\"urich, Switzerland}
\author{P.~Crivelli}\affiliation{\it ETH Z\"urich, Institute for Particle Physics and Astrophysics, CH-8093 Z\"urich, Switzerland}
\author{E.~Depero}\affiliation{\it ETH Z\"urich, Institute for Particle Physics and Astrophysics, CH-8093 Z\"urich, Switzerland}
\author{A.~V.~Dermenev}\affiliation{\it Institute for Nuclear Research, 117312 Moscow, Russia}
\author{S.~V.~Donskov}\affiliation{\it State Scientific Center of the Russian Federation Institute for High Energy Physics of National Research Center 'Kurchatov Institute' (IHEP), 142281 Protvino, Russia}
\author{R.~R.~Dusaev}\affiliation{\it Tomsk State Pedagogical University, 634061 Tomsk, Russia}
\author{T.~Enik}\affiliation{\it  Joint Institute for Nuclear Research, 141980 Dubna, Russia}
\author{N.~Charitonidis}\affiliation{\it CERN, European Organization for Nuclear Research, CH-1211 Geneva, Switzerland}
\author{A.~Feshchenko}\affiliation{\it  Joint Institute for Nuclear Research, 141980 Dubna, Russia}
\author{V.~N.~Frolov}\affiliation{\it  Joint Institute for Nuclear Research, 141980 Dubna, Russia}
\author{A.~Gardikiotis}\affiliation{\it Physics Department, University of Patras, Patras, Greece}
\author{S.~G.~Gerassimov }\affiliation{\it P.N. Lebedev Physics Institute, Moscow, Russia, 119 991 Moscow, Russia}\affiliation{\it Technische Universit\"at
M\"unchen, Physik Dept., 85748 Garching, Germany}
\author{S.~N.~Gninenko$^{\ast}$\footnote{Corresponding author, Sergei.Gninenko@cern.ch}}\affiliation{\it Institute for Nuclear Research, 117312 Moscow, Russia}
\author{M.~H\"osgen}\affiliation{\it Universit\"at Bonn, Helmholtz-Institut f\"ur Strahlen-und Kernphysik, 53115 Bonn, Germany}
\author{M.~Jeckel}\affiliation{\it CERN, European Organization for Nuclear Research, CH-1211 Geneva, Switzerland}
\author{A.~E.~Karneyeu}\affiliation{\it Institute for Nuclear Research, 117312 Moscow, Russia}
\author{G.~Kekelidze}\affiliation{\it  Joint Institute for Nuclear Research, 141980 Dubna, Russia}
\author{B.~Ketzer}\affiliation{\it Universit\"at Bonn, Helmholtz-Institut f\"ur Strahlen-und Kernphysik, 53115 Bonn, Germany}
\author{D.~V.~Kirpichnikov}\affiliation{\it Institute for Nuclear Research, 117312 Moscow, Russia}
\author{M.~M.~Kirsanov}\affiliation{\it Institute for Nuclear Research, 117312 Moscow, Russia}
\author{I.~V.~Konorov}\affiliation{\it P.N. Lebedev Physics Institute, Moscow, Russia, 119 991 Moscow, Russia} \affiliation{\it Technische Universit\"at
M\"unchen, Physik Dept., 85748 Garching, Germany}
\author{S.~G.~Kovalenko}\affiliation{\it Universidad T\'{e}cnica Federico Santa Mar\'{i}a, 2390123 Valpara\'{i}so, Chile}
\author{V.~A.~Kramarenko}\affiliation{\it  Joint Institute for Nuclear Research, 141980 Dubna, Russia}\affiliation{\it Skobeltsyn Institute of Nuclear Physics, Lomonosov Moscow State University, Moscow, Russia}
\author{L.~V.~Kravchuk}\affiliation{\it Institute for Nuclear Research, 117312 Moscow, Russia}
\author{ N.~V.~Krasnikov}\affiliation{\it Institute for Nuclear Research, 117312 Moscow, Russia}
\author{S.~V.~Kuleshov}\affiliation{\it Universidad T\'{e}cnica Federico Santa Mar\'{i}a, 2390123 Valpara\'{i}so, Chile}
\author{V.~E.~Lyubovitskij}\affiliation{\it Tomsk State Pedagogical University, 634061 Tomsk, Russia}\affiliation{\it Universidad T\'{e}cnica Federico Santa Mar\'{i}a, 2390123 Valpara\'{i}so, Chile}
\author{V.~Lysan}\affiliation{\it  Joint Institute for Nuclear Research, 141980 Dubna, Russia}
\author{V.~A.~Matveev}\affiliation{\it  Joint Institute for Nuclear Research, 141980 Dubna, Russia}
\author{Yu.~V.~Mikhailov}\affiliation{\it State Scientific Center of the Russian Federation Institute for High Energy Physics of National Research Center 'Kurchatov Institute' (IHEP), 142281 Protvino, Russia}
\author{D.~V.~Peshekhonov}\affiliation{\it  Joint Institute for Nuclear Research, 141980 Dubna, Russia}
\author{V.~A.~Polyakov}\affiliation{\it State Scientific Center of the Russian Federation Institute for High Energy Physics of National Research Center 'Kurchatov Institute' (IHEP), 142281 Protvino, Russia}
\author{B.~Radics}\affiliation{\it ETH Z\"urich, Institute for Particle Physics and Astrophysics, CH-8093 Z\"urich, Switzerland}
\author{R.~Rojas}\affiliation{\it Universidad T\'{e}cnica Federico Santa Mar\'{i}a, 2390123 Valpara\'{i}so, Chile}
\author{A.~Rubbia}\affiliation{\it ETH Z\"urich, Institute for Particle Physics and Astrophysics, CH-8093 Z\"urich, Switzerland}
\author{V.~D.~Samoylenko}\affiliation{\it State Scientific Center of the Russian Federation Institute for High Energy Physics of National Research Center 'Kurchatov Institute' (IHEP), 142281 Protvino, Russia}
\author{V.~O.~Tikhomirov}\affiliation{\it P.N. Lebedev Physics Institute, Moscow, Russia, 119 991 Moscow, Russia}
\author{D.~A.~Tlisov}\affiliation{\it Institute for Nuclear Research, 117312 Moscow, Russia} 
\author{A.~N.~Toropin}\affiliation{\it Institute for Nuclear Research, 117312 Moscow, Russia}
\author{A.~Yu.~Trifonov}\affiliation{\it Tomsk State Pedagogical University, 634061 Tomsk, Russia}
\author{B.~I.~Vasilishin}\affiliation{\it Tomsk State Pedagogical University, 634061 Tomsk, Russia}
\author{G.~Vasquez Arenas}\affiliation{\it Universidad T\'{e}cnica Federico Santa Mar\'{i}a, 2390123 Valpara\'{i}so, Chile}
\author{P.~V.~Volkov}\affiliation{\it  Joint Institute for Nuclear Research, 141980 Dubna, Russia}\affiliation{\it Skobeltsyn Institute of Nuclear Physics, Lomonosov Moscow State University, Moscow, Russia}
\author{V.~Volkov}\affiliation{\it Skobeltsyn Institute of Nuclear Physics, Lomonosov Moscow State University, Moscow, Russia}
\author{P.~Ulloa}\affiliation{\it Universidad T\'{e}cnica Federico Santa Mar\'{i}a, 2390123 Valpara\'{i}so, Chile}

%
%
\collaboration{The NA64 Collaboration}\noaffiliation
\vskip 0.25cm